\documentclass[reprint,groupedaddress,amsmath,amssymb,aps,prx,floatfix,longbibliography]{revtex4-1}

\usepackage{amsmath,amssymb,graphicx,mathrsfs,amsfonts}

\usepackage{color}
\draft 

\begin{document}

\title{Frequency measurements of superradiance from the strontium clock transition} 



\author{Matthew A. Norcia}
\affiliation{JILA, NIST, and University of Colorado, 440 UCB, Boulder, CO  80309, USA}
\author{Julia R. K. Cline}
\affiliation{JILA, NIST, and University of Colorado, 440 UCB, Boulder, CO  80309, USA}
\author{Juan A. Muniz}
\affiliation{JILA, NIST, and University of Colorado, 440 UCB, Boulder, CO  80309, USA}
\author{John M. Robinson}
\affiliation{JILA, NIST, and University of Colorado, 440 UCB, Boulder, CO  80309, USA}
\author{Ross B. Hutson}
\affiliation{JILA, NIST, and University of Colorado, 440 UCB, Boulder, CO  80309, USA}
\author{Akihisa Goban}
\affiliation{JILA, NIST, and University of Colorado, 440 UCB, Boulder, CO  80309, USA}
\author{G. Edward Marti}
\affiliation{JILA, NIST, and University of Colorado, 440 UCB, Boulder, CO  80309, USA}
\author{Jun Ye}
\affiliation{JILA, NIST, and University of Colorado, 440 UCB, 
Boulder, CO  80309, USA}
\author{James K. Thompson}
\affiliation{JILA, NIST, and University of Colorado, 440 UCB, 
Boulder, CO  80309, USA}
\email[]{matthew.norcia@colorado.edu}

\date{\today}

\begin{abstract}

We present the first characterization of the spectral properties of superradiant light emitted from the ultra-narrow, 1~mHz linewidth optical clock transition in an ensemble of cold $^{87}$Sr atoms.  Such a light source has been proposed as a next-generation active atomic frequency reference, with the potential to enable high-precision optical frequency references to be used outside laboratory environments.  By comparing the frequency of our superradiant source to that of a state-of-the-art cavity-stabilized laser and optical lattice clock, we observe a fractional Allan deviation of $6.7(1)\times 10^{-16}$ at 1 second of averaging, establish absolute accuracy at the 2~Hz ($4\times 10^{-15}$ fractional frequency) level, and demonstrate insensitivity to key environmental perturbations.  
\end{abstract}


\maketitle 
\section{Introduction}\label{sec:intro}
The  development of atomic clocks has lead to a wealth of applications from technology to fundamental physics.  Atomic clocks are currently the most precise and accurate absolute frequency references \cite{Bloom2014,Schioppo2017,campbell2017fermi,nicholson2015systematic,chou2010frequency,LudlowRev2015} and are used to define the second \cite{gill2005optical,gill2011should,CIPMrecommendation2015}.  Precision timekeeping enables the synchronization needed for communication and navigation systems \cite{grewal2013global,major2007quantum}, even through direct optical links across countries \cite{lisdat2016clock}, and for very-long-baseline radio telescopes \cite{doeleman2011adapting}. Precision frequency metrology has been proposed as a means of studying quantum many-body physics \cite{martin2013quantum,Zhang19092014}, searching for exotic physics \cite{derevianko2014hunting,Derevianko2016,safronova2017search} and exploring fundamental quantum limits imposed by gravity \cite{zych2011quantum,pikovski2015universal,ruiz2017entanglement}.  

Atomic clocks typically function by combining a stable oscillator (e.g. a crystal oscillator or laser) and an atomic frequency reference, which is an atom or collection of atoms with a well-suited transition between internal states (the ``clock transition") to which the oscillator is stabilized.  The oscillator provides short-term stability, while the atomic frequency reference provides long-term stability and absolute accuracy.  

\begin{figure}[!htb]
\includegraphics[width=3.375in]{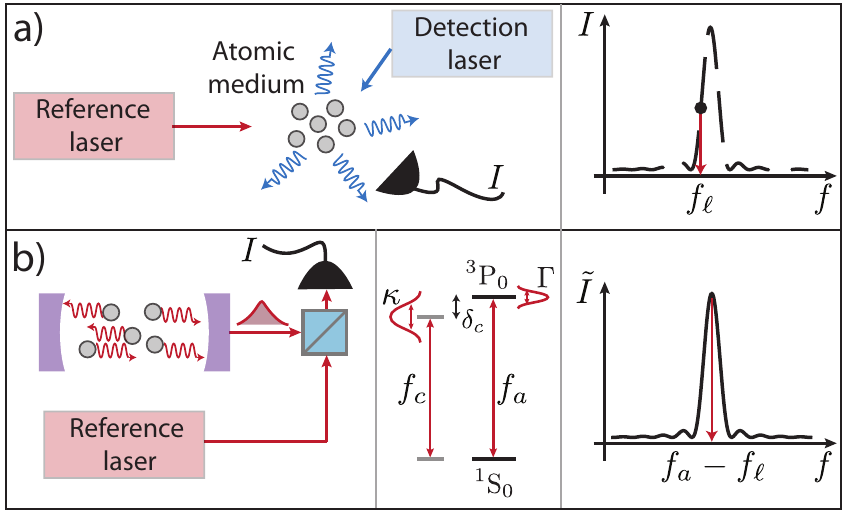}
\caption{Passive and active optical frequency references. (a) In a passive optical frequency reference, the ``clock transition" in the atomic medium is driven by radiation from a laser at frequency $f_\ell$.  The laser frequency is then determined relative to the clock transition frequency through a population measurement of the two clock states using a fluorescence signal ($I$) obtained from a separate atomic transition.  The fluorescence signal provides information about the atomic response at frequency $f_\ell$, which can be used to stabilize the frequency of the laser.  (b) In an active optical frequency reference, light is collected directly from the atomic clock transition.  This light can be compared to the reference laser by forming a heterodyne beat-note.  The power spectral density of the resulting signal ($\tilde{I}$) provides information about the atomic system at all offset frequencies from $f_\ell$ within a large detection bandwidth set by the detection electronics.  In our system, we use  $^{87}$Sr atoms emitting into a high-finesse optical cavity on a roughly 1~mHz linewidth clock transition as the active atomic medium. The clock transition frequency, $f_a$, is nearly equal to a cavity resonance frequency, $f_c$, up to a tunable offset $\delta_c = f_a-f_c$.  The cavity linewidth $\kappa = 2\pi\times 145$~kHz greatly exceeds the atomic linewidth, placing the system in a highly damped, ``superradiant" regime.  }
\label{fig:exp}
\end{figure}

The atomic frequency reference can either be active or passive \cite{riehle2006frequency}. In a passive reference, as sketched in Fig.~\ref{fig:exp}(a), radiation from the oscillator is applied to the atoms in a spectroscopic sequence that maps the oscillator's frequency to internal state populations in the atoms.  The populations are then measured to infer and stabilize the oscillator's frequency \cite{LudlowRev2015,Poli2014,derevianko2011colloquium}. Examples include caesium and rubidium clocks in the microwave domain \cite{Heavner14,Guena14}, and strontium, ytterbium, and aluminum ion clocks in the optical domain \cite{LudlowRev2015}. 

In an active reference, radiation is collected directly from the clock transition, as depicted in Fig.~\ref{fig:exp}(b). While passive frequency references have so far demonstrated superior long-term stability and absolute accuracy compared to their active counterparts, active references can have substantial advantages in terms of detection bandwidth and dynamic range, which can be especially important in situations where the system performance is limited by large excursions of the oscillator frequency (discussed further in Section~\ref{sec:system}). Hydrogen masers are currently used as active frequency references in the microwave domain to provide complimentary short-term stability to microwave atomic clocks \cite{Goldenberg1960,riehle2006frequency}.  Until now, there have been no high-precision active atomic frequency references in the optical domain. 

Operating a frequency reference in the optical domain instead of the microwave domain provides a huge fundamental advantage that has allowed passive optical frequency references to overtake their microwave counterparts in both precision and accuracy \cite{LudlowRev2015,Poli2014,MEH10,MYC09}.  Here, we demonstrate the first active optical frequency reference to realize this advantage, achieving a measured fractional frequency stability at short times surpassing that of hydrogen masers by roughly two orders of magnitude.    


\section{Experimental system}\label{sec:system}

Our system, also described in \cite{norcia2016superradiance}, consists of an ensemble of up to several $10^5$ $^{87}$Sr atoms confined within a high finesse optical cavity by a near-magic wavelength \cite{Yemagictrap2008} optical lattice that is supported by the cavity. This lattice confines the atoms tightly along the cavity axis (in the so-called Lamb-Dicke regime) so as to eliminate first-order Doppler shifts, while imparting near equal shifts to the ground and excited states of the lasing transition. The cavity is locked at frequency $f_c$, close to the frequency $f_a$ of the dipole-forbidden optical clock transition between the ground state $^1$S$_0$ and the excited state $^3$P$_0$, as shown in Fig.~\ref{fig:exp}(b).  The wavelength of the clock transition is near 698~nm, and it has a radiative linewidth of 1.1(3)~mHz \cite{Ye2007}.  The cavity detuning from the clock transition is $\delta_c = f_c-f_a$. At 698~nm, the cavity's finesse is $2.5\times 10^4$, and its linewidth is $\kappa = 2\pi\times 145$~kHz. In typical operation, $\delta_c$ is nominally zero ($\delta_c \ll \kappa$), but can be varied for characterization purposes.  


We prepare a superradiant state by initializing the atoms in one or more ground states, and then applying a coherent drive to the clock transition using a laser that is coupled through the optical cavity (we will call this the state preparation pulse). Both the state preparation pulse and the optical cavity are nominally on resonance with the clock transition. This prepares each atom in a superposition of the ground and excited states with the correct relative phase between atoms to radiate collectively into the cavity mode, and a population inversion set by the pulse duration. A pulse that prepares a maximally inverted state lasts about 100~$\mu$s.  As soon as the state preparation pulse is turned off, the atoms emit collectively into the cavity. We refer to this process as superradiant emission, which describes the collective radiation of an ensemble of emitters into a highly damped optical mode.   After the atoms emit a superradiant pulse, a new ensemble of atoms is laser cooled and trapped and the sequence is repeated with a typical total cycle time $T_c = 1.1$~s.

In order to study the spectral properties of the superradiant light, we form a heterodyne beat-note between the light emitted from the strontium atoms and light from a state-of-the-art stable laser system that we refer to as the reference laser, as shown in Fig.~\ref{fig:exp}(b). The reference laser system is described in \cite{Swallows2012,Bishof_2013} and is stabilized to an optical reference cavity with a thermal noise floor of $1\times 10^{-16}$ from 1 to 1000~seconds, and a linewidth of 26~mHz. 
Further details on the experimental setup and data processing can be found in Appendix I.


\begin{figure}[!htb]
\includegraphics[width=3.375in]{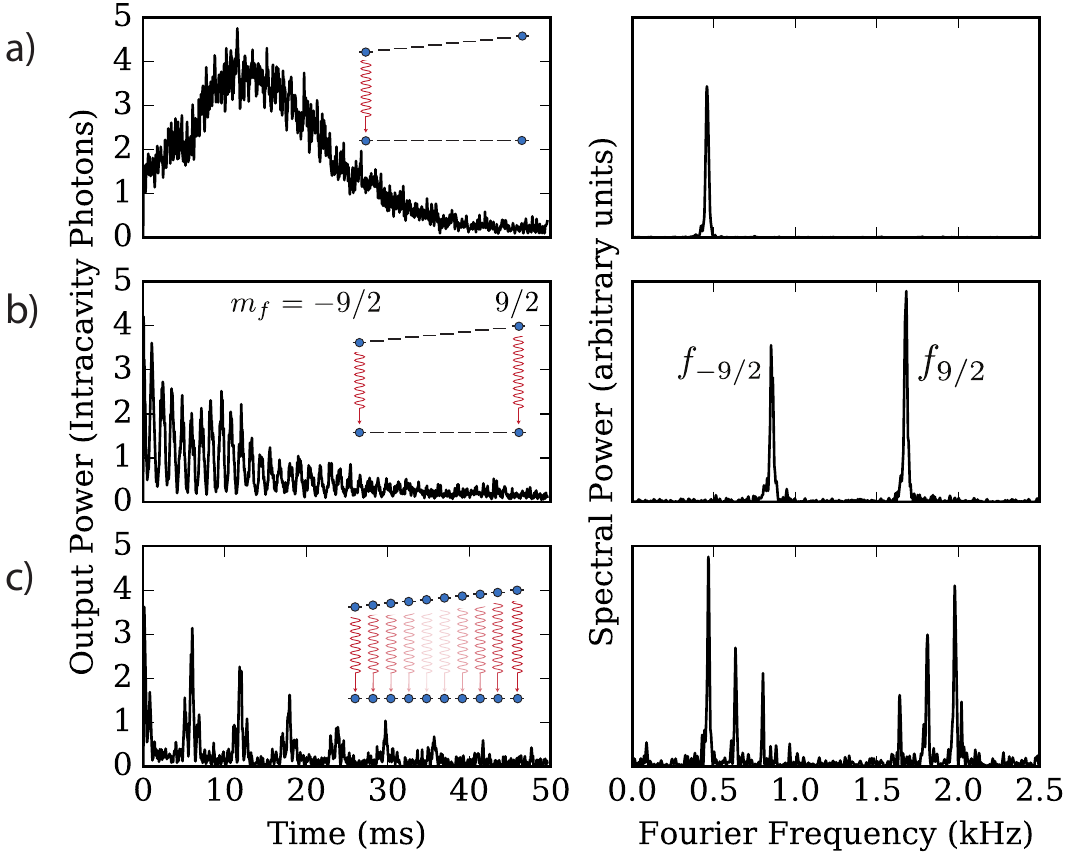}
\caption{Superradiant pulses in the time and frequency domains with different initial nuclear spin state populations. Atomic state population is initialized (a) with all atoms in a single stretched state, (b) with atoms in both stretched states, and (c) with all ten spin states populated. Columns represent the output power measured in the time domain (left), the initial spin state population (inset), and power spectrum of output power measured relative to stable reference laser up to an arbitrary frequency offset (right), for single trials of the experiment. In the frequency domain, we observe a single peak in the power spectrum when we populate only one stretched state, two peaks when we populate both stretched states, and many peaks when we populate all ten Zeeman sublevels.}
\label{fig:sr}
\end{figure}

Fig.~\ref{fig:sr} shows characteristic superradiant pulses in the time and frequency domain, for different initial nuclear spin state populations.  The ground and excited states of the lasing transition in $^{87}$Sr each have ten Zeeman sublevels ($F=9/2$), which are non-degenerate in the presence of an applied magnetic field.  Because the state preparation pulse is polarized along the magnetic field, it maintains the initial spin projection $m_f$ of the ground state.  Further, because the state preparation pulse establishes coherence between the ground and excited states, the radiated light is also polarized along the magnetic field and we do not observe lasing between different values of $m_f$.  

In row (a) of Fig.~\ref{fig:sr}, we optically pump the atoms into one stretched state ($|m_f| = 9/2$) prior to the state preparation pulse. If we measure the output power versus time, we observe the immediate onset of superradiant emission that lasts a few 10's of ms. The power spectral density (PSD) of the beat-note between the superradiant light and the reference laser shows a single peak with width of order 10~Hz, corresponding to the finite duration of the pulse.

\begin{figure*}[!htb]
\includegraphics[width=6.75in]{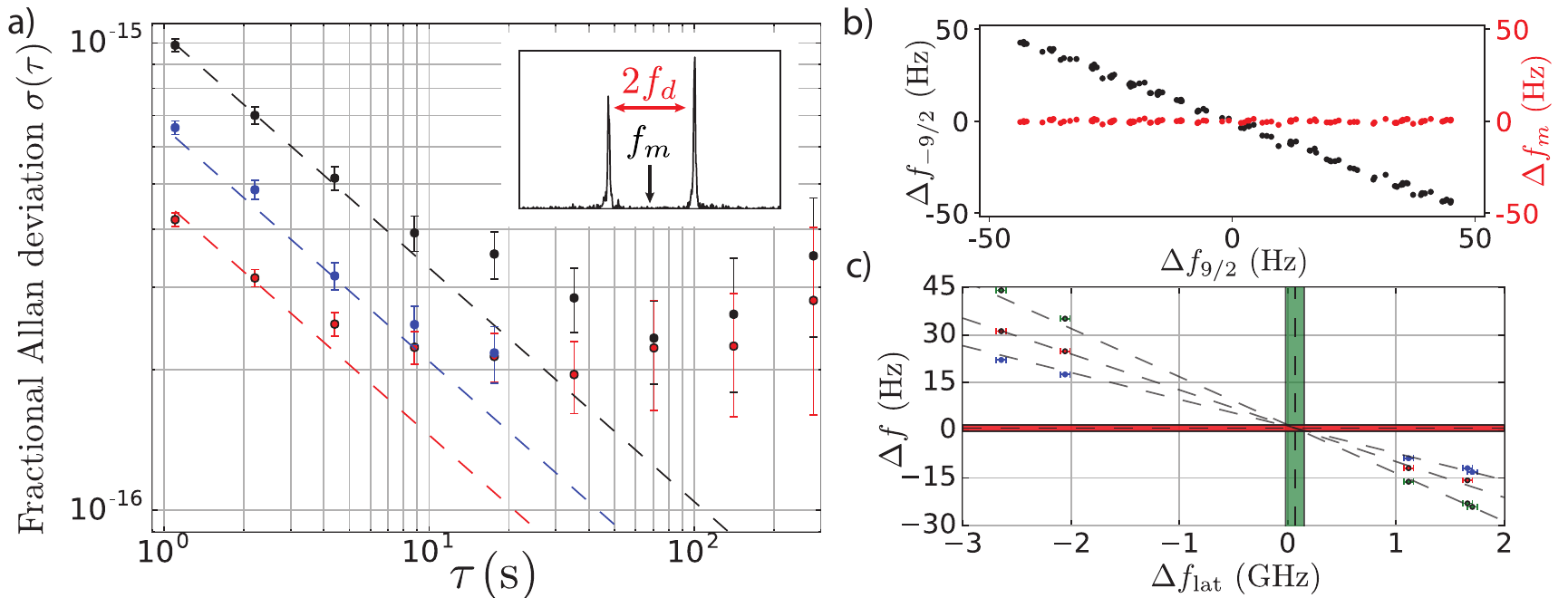}
\caption{Stability and accuracy of the superradiant laser. (a) Short-term stability of superradiant light source expressed as fractional Allan deviation $\sigma(\tau)$. We populate both stretched states and compute the Allan deviation for two quantities: the average frequency of the two lasing transitions $f_m$ (black, blue points) and half the difference of the two lasing transition frequencies $f_d$ (red points). During a longer dataset, we measure a short-term fractional instability of $\sigma(\tau) = 1.04(4)\times 10^{-15}/\sqrt{\tau/s}$ for the average frequency (black points), and a value of $\sigma(\tau) = 6.7(1)\times 10^{-16}/\sqrt{\tau/s}$ for a shorter data set taken under particularly quiet conditions (blue points).  (b) Real-time insensitivity to changing magnetic fields is obtained by simultaneously fitting the frequencies of two radiating transition.  Perturbations to the two individual transitions $\Delta f_{\pm 9/2}$ (black points) are suppressed to below the half-percent level in the sum frequency $\Delta f_m$ (red points).  
(c) Lattice-induced frequency shifts. Frequency difference between the superradiant laser and an optical lattice clock for different lattice laser detunings and power levels. Different colored points correspond to different lattice depths: $U = 1025, \ 760, \ 570 \  \mathrm{E_{rec}}$ for green, red and blue points respectively, where $\mathrm{E_{rec}}$ is the lattice recoil energy. The horizontal axis shows $\Delta f_\mathrm{lat} = f_\mathrm{lat}-f_\mathrm{magic}^\mathrm{wm}$, where $f_\mathrm{magic}^\mathrm{wm}$ is the expected value for the magic wavelength as measured by our wavemeter. A simultaneous fit (see text) to all data points returns a nominal frequency offset consistent with zero, represented by the horizontal dashed line, with the uncertainty of the fit represented by the red band. The vertical dashed line represents the fit magic wavelength, relative to $f_\mathrm{magic}^\mathrm{wm}$, with fit uncertainty given by the green band.}
\label{fig:allan}
\end{figure*}

Instead of populating a single spin state, we may populate both stretched states simultaneously and observe lasing on both transitions at the same time, as shown in Fig.~\ref{fig:sr}(b). In this configuration, we are preparing a spin mixture, in contrast to a coherent superposition between the two stretched states --- half of the atoms are in one stretched state, and the other half are in the other.  Because the Rabi frequency of the state preparation pulse exceeds the Zeeman splitting between the two transitions, both experience a similar initial rotation.  In the time domain, we see a beating at the frequency of the Zeeman splitting between the two transitions. In the beat-note's PSD, we see two distinct peaks at frequencies that we label $f_{9/2}$ and $f_{-9/2}$, corresponding to light emitted from the two stretched transitions. Because these two peaks shift in equal and opposite directions in response to magnetic fields, this configuration can be used to derive a magnetic-field insensitive signal, as described in Fig.~\ref{fig:allan}(b) and the subsequent discussion.  

Finally, we can populate all ten nuclear spin states, and observe interference between the fields radiated by atoms populating the ten nuclear spin states as in Fig.~\ref{fig:sr}(c). This manifests as a series of short bursts of power as the ten transitions come into phase with one-another, separated by dark periods when their radiated fields cancel. In the beat-note's PSD, we see peaks associated with different transitions, with heights that reflect the relative strengths of the Clebsh-Gordan coefficients of the transitions. 

The data shown in each panel of Fig.~\ref{fig:sr} represent a single run of the experiment.  In contrast to a passive clock, where one would have to run the experiment at least once for every point on the frequency axis in order to gain information about the atomic response at that frequency, we get all of this information from a single pulse.  For example, resolving the ten nuclear spin transitions of Fig.~\ref{fig:sr}(c) in a passive clock would require a scan of the clock laser frequency, as in \cite{Boydscience2006}. 

Furthermore, in a standard passive clock, dynamic range and precision are fundamentally linked -- one only gains useful information if the frequency of the local oscillator is within the bandwidth of the measurement ($1/T$, where $T$ is the interrogation time) from the atomic transition.  $T$ can be reduced in order to increase the bandwidth, but this also increases the width of the spectroscopic feature, degrading precision.  Large frequency excursions are thus difficult to measure with high precision in a passive clock.  With an active source however, as long as the cavity that surrounds the atoms is within a few cavity linewidths of the atomic transition (several hundred kHz in our case), the output of the active clock provides a frequency reference with no reduction in precision.  

The large dynamic range of the active reference may prove valuable for the operation of high precision optical frequency references outside of laboratory environments, and in mobile applications.  In these contexts, vibrations and temperature fluctuations could cause large and rapid perturbations to the frequency of the local oscillator laser which would be very difficult to suppress using a passive atomic frequency reference.  The large bandwidth and dynamic range of the active reference could provide substantial advantages in these conditions.

\section{Stability and accuracy}\label{sec:stability}

To quantify the stability of our superradiant source, we rely on the commonly-used metric of the fractional Allan deviation \cite{Allan66}, which is defined as $\sigma(\tau) = \sqrt{\frac{1}{2f^2_a} \left\langle \left( \bar{f}_{n+1} - \bar{f}_{n}\right) ^2\right\rangle}$, where each $\bar{f}_n$ is the measured frequency averaged over the n-th time interval of duration $\tau$, and $f_a$ is the atomic transition frequency. In each run of the experiment, the frequency is determined from the center of a Lorentzian fitted to one or more peaks in the power spectrum of the beat-note.  For $\tau > T_c$, each $\bar{f}_n$ is obtained by averaging the fitted centers of the Fourier peaks obtained in several subsequent trials.  

The process of extracting the center frequency from a fit suppresses the effects of frequency variations that occur on timescales short compared to the duration of the superradiant pulse.  As such, our reported Allan deviations should not be extrapolated to measurement durations shorter than the cycle time $T_c$.  For $\tau \geq T_c$ however, where we are interested in average frequencies over long periods of time, this filtering is unimportant and our reported Allan deviations can be compared directly to those calculated using other measurement protocols.  The impact of the pulse duration being shorter than the cycle time is discussed in Appendix III.

The Allan deviation of the frequencies of our superradiant pulses, $\sigma(\tau)$, is shown in Fig.~\ref{fig:allan}(a), after subtracting a linear drift associated with the reference laser. In this measurement, we populate both stretched states, as in Fig.~\ref{fig:sr}(b) and compute the Allan deviation for two quantities: the average frequency of the two lasing transitions $f_m= (f_{9/2} + f_{-9/2})/2$ (black, blue points), which provides an indictation of the stability of our system if used as an optical frequency reference, and half the difference of the two lasing transition frequencies $f_d= (f_{9/2} - f_{-9/2})/2$ (red points), which provides a means of assessing the fundamental limits of our system. For the black points, $\sigma(\tau) = 1.04(4) \times 10^{-15}/\sqrt{\tau/s}$ for small $\tau$. Our most favorable measurement of this value (blue points) was $\sigma(\tau) = 6.7(1) \times 10^{-16}/\sqrt{\tau/s}$, obtained for a different, shorter set of data.  This corresponds to a standard deviation of 300~mHz. 

For typical hydrogen masers, the short term stability is less than $ 10^{-12}$ at 1~second of averaging, while cryogenic hydrogen masers can achieve stabilities lower than $ 10^{-13}$ at 1 second \cite{riehle2006frequency,Weiss1996}. The recently-obtained record for short-term stability in an atomic frequency reference is $6 \times 10^{-17}/\sqrt{\tau/s}$, achieved by interleaved interrogation of two ensembles of Ytterbium at NIST \cite{Schioppo2017}.  Our value is encouragingly close, given the maturity and careful engineering of current optical lattice clocks.  

The stability of $f_m$ is fundamentally bounded by photon shot noise.  In order to determine how close we are to this limit, we consider the short-term stability of the difference frequency $f_d$ (red points of Fig.~\ref{fig:allan}(a)).  This constitutes a synchronous comparison of the atoms populating each of the two stretched states, similar to synchronous comparisons used to extract the quantum-limited stability of passive optical lattice clocks in the presence of local oscillator noise \cite{nicholson2012comparison, Schioppo2017, campbell2017fermi}.  

The frequency difference $f_d$ is insensitive to perturbations that affect the measurement of both lasing transitions in the same manner -- for example noise on the reference laser, cavity pulling effects, and shifts associated with atomic density. For the data set corresponding to the black points of Fig.~\ref{fig:allan}(a), $f_d$ exhibits a short-term stability of $\sigma(\tau) = 4.5(2) \times 10^{-16}/\sqrt{\tau/s}$. This result can be compared to a recent differential measurement between two regions of a strontium Fermi-degenerate three-dimensional optical lattice clock that obtained a fractional Allan deviation of $3.1 \times 10^{-17}/\sqrt{\tau/s}$ \cite{campbell2017fermi}.

In these passive clocks, atomic quantum projection noise dominates over shot noise associated with the photons used to measure atomic state populations, as many photons can be scattered and collected from each atom in the fluorescence detection process.  In contrast, in our active system each atom currently emits only a single photon per measurement. Because our detection efficiency ($\eta = 0.06$, see Appendix I and III) is significantly lower than unity, photon shot noise is a larger source of noise than atomic projection noise. By deliberately attenuating the output of our active reference, we have verified that the short term-instability in $f_d$ is dominated by photon shot noise. 

Much of the difference between the measured Allan deviations for $f_m$ and $f_d$ can be attributed to noise on the reference laser, which is exacerbated by the fact that the duration of our pulses is shorter than the cycle time of the experiment \cite{Dick1987} (see Appendix III for details).  Other likely contributions include fluctuations in atom number and cavity frequency.  


For longer averaging windows (large $\tau$), the measured Allan deviations flatten out. The reference laser has a noise floor due to thermal perturbations of the length of the reference cavity at around $\sigma(\tau) = 1\times 10^{-16}$ from 1 to 1000 seconds \cite{Swallows2012,Bishof_2013}, which contributes to the floor of the average frequencies $f_m$ (black points). The Allan deviation for the difference frequencies, $f_d$, (red points) is insensitive to noise on the reference cavity. These points also hit a floor at $\sigma(\tau) \simeq 2\times10^{-16}$, perhaps due to slowly varying magnetic fields in the lab.  To our knowledge, the fact that the asymptotic values of these two quantities is so similar is a coincidence.  

In Fig~\ref{fig:allan}(b), we demonstrate the insensitivity of $f_m$ to magnetic fields. The clock transition frequencies $f_{\pm 9/2}$ have a low but finite magnetic field sensitivity of $108.4\times m_f$ (Hz/G)  \cite{boyd2007nuclear}, which is opposite in sign for the two transitions. 
For demonstration, we deliberately add a randomized magnetic field between different runs of the experiment.  While this leads to large shifts of the individual peaks, $\Delta f_{9/2}$ and $\Delta f_{-9/2}$, the shift to $f_m$, called $\Delta f_m$, is suppressed by at least two orders of magnitude over our range of applied magnetic fields. In contrast to passive clocks, where the spin polarization is alternated between experimental trials to achieve a magnetic field insensitive signal, our method provides real-time rejection of magnetic field noise, making it effective at eliminating errors from time-varying magnetic fields.  

In order to assess the absolute accuracy of the superradiant light source, we perform a comparison with the optical lattice clock described in \cite{campbell2017fermi}. For this comparison, we account for known shifts to the frequencies of both the lattice clock and superradiant system at the Hz level, as detailed in Appendix II. 

The largest such shift is due to the differential AC Stark shift from the optical lattice. Because the cavity must be on resonance with the clock transition at 698~nm and the lattice frequency must be on resonance with another mode of the cavity near 813~nm, we do not have complete control of the lattice frequency.  This prevents us from operating at exactly the magic wavelength, at which the shift to the clock transition vanishes. The closest longitudinal TEM$_{00}$ mode to the magic wavelength may be up to half a free spectral range (1.85~GHz) away.  

To account for this, we take data with the lattice at four frequencies near the magic wavelength and at three lattice depths at each of these frequencies (Fig. \ref{fig:allan}(c)). We then perform a fit of the form:
$\Delta f = a(f_\mathrm{lat} - f_\mathrm{magic})\times U + f_{\mathrm{offset}}$
to all of the data points. Here, $f_\mathrm{lat}$ and $U$ are the measured lattice frequencies and trap depths, and
$a$, $f_\mathrm{magic}$, and $f_{\mathrm{offset}}$ are fit parameters that represent the sensitivity of the clock transition to lattice detuning, the magic wavelength frequency, and the systematic frequency offset between the superradiant laser and the optical lattice clock, which is our primary quantity of interest. 

After accounting for other known shifts, dominated by atomic collisions (see Appendix II), the measured frequency difference between our superradiant light and the value measured by the optical lattice clock is $1 \pm 2$~Hz (fractionally $2(4)\times 10^{-15}$).  


\begin{figure}[!htb]
\includegraphics[width=3.375in]{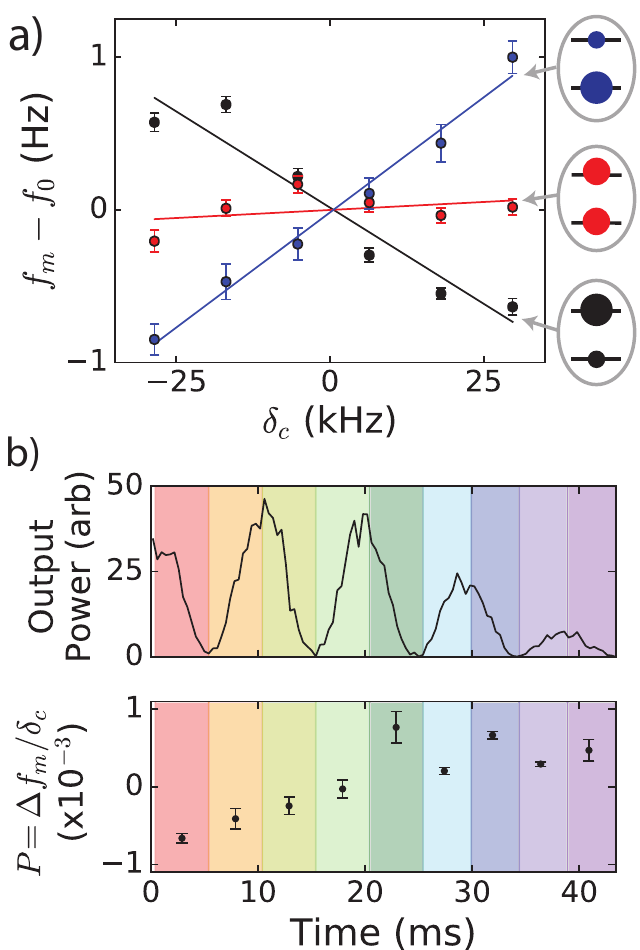}
\caption{(a) Sensitivity of superradiant light frequency, $f_m$, to detuning of the cavity from resonance, $\delta_c$, for different initial population inversions, as sketched on the side insets. An offset frequency $f_0$ set by the detuning of the reference laser from the strontium transition is subtracted for each data set. When tuned to minimize sensitivity to cavity detuning (red line, points) we observe a pulling coefficient $P = \Delta f_m/\delta_c = 2(3) \times 10^{-6}$. Changing the duration of the state preparation pulse leads to an increased absolute value of $P$ (black, blue points).  (b) Atomic inversion changes throughout the superradiant pulse, leading to a chirp in the frequency of the emitted light for finite cavity detuning. The emitted power is shown in the upper panel for the first 45~ms of the pulse. The pulling coefficient at different time intervals is computed from frequency measurements made on subsets of the pulse (as indicated by colored vertical bands) for the red data set shown in (a). The pulling coefficient of the entire pulse can be made insensitive to cavity detuning by choosing an initial inversion for which the time-averaged (weighted by instantaneous power) frequency shift is zero. }
\label{fig:fig4}
\end{figure}

\section{Cavity Pulling}\label{sec:pulling}

Next, we characterize the sensitivity of the frequency of the superradiant light to changes in the cavity resonance frequency. Long-range exchange interactions mediated by the optical cavity give rise to a frequency shift that depends dispersively on cavity detuning, and linearly on atomic population inversion \cite{Norcia2017, hu2017vacuum}. These interactions lead to so-called one-axis-twisting dynamics, and a many-body energy gap. The key effect of these dynamics for our purposes here is that small perturbations of the cavity detuning from atomic resonance, $\delta_c$, cause the frequency of the laser to be ``pulled" by an amount $P \delta_c$, relative to its value when the cavity is on resonance.  We refer to $P$ as the pulling coefficient. While the cavity detuning is normally stabilized near zero $|\delta_c| < 5$~kHz $ \ll\kappa$ using piezoelectric actuators (PZTs) on the cavity mirrors, we next use the PZTs to introduce a finite and controllable detuning.  

Because the pulling coefficient is proportional to the population inversion, it changes signs as the inversion changes from positive to negative during a superradiant pulse.  When we fit the center frequency of a Fourier peak, we are sensitive to a value of $P$ that is averaged over different values of inversion. By preparing states for which the average inversion during the superradiant pulse is near zero, we can make the average $P$ very small.  


In Fig~\ref{fig:fig4}(a), we vary the duration of the state preparation pulse in order to prepare states with differing initial inversion, and measure the frequency shift that results from a shift in cavity frequency. We can tune the duration of the state preparation pulse to give a low pulling coefficient of $P = 2(3) \times 10^{-6}$ (red points). Pulse durations 13\% longer and 7\% shorter yield pulling coefficients roughly an order of magnitude higher ($P= -2.5(5)\times 10^{-5}$ and $P = 3.0(3) \times 10^{-5}$ for the black and blue points, respectively).  For the data used to measure stability and accuracy, we tuned the initial inversion near the point of minimum sensitivity to cavity frequency deviations.  Further, we nominally tuned the cavity on resonance with the atomic transition ($\delta_c = 0$) for each point in Fig.~\ref{fig:allan}c.  As such, we do not expect cavity pulling to have contributed a systematic shift to our accuracy assessment.  

We directly observe that as the inversion changes during a superradiant pulse, the frequency of the output light indeed chirps in time. A typical single-shot measurement of optical power versus time is shown in the upper panel of Fig.~\ref{fig:fig4}(b), with both stretched states populated as in Fig.~\ref{fig:sr}(b). In the lower panel of Fig.~\ref{fig:fig4}(b), we calculate the pulling coefficient from 5~ms long subsets of the data corresponding to the red points of Fig.~\ref{fig:fig4}(a). The pulling coefficient switches sign during the pulse as inversion goes from positive to negative.  
The average of the pulling coefficient measurements (weighted by the output power) is consistent with the pulling coefficient for the entire pulse.  

For any of these configurations, the output frequency of the laser is highly insensitive to the cavity frequency. This is a key promise of superradiant lasers -- because the output frequency is determined by the atomic transition frequency rather than the cavity resonance frequency, a superradiant laser is largely insensitive to technical and fundamental thermal fluctuations in cavity length that limit today's most stable lasers \cite{Braginsky1999,Levin1998,Cole2013,Matei2017,Zhang2017}.

It is important to note that minimizing $P$ by carefully setting the initial inversion is only possible in pulsed operation.  In steady-state, the existence of an equilibrium condition requires positive population inversion, and thus leads to a finite pulling coefficient \cite{PhysRevLett.72.3815, bohnet2012steady, norcia2015cold}. For our system under realistic steady-state operating conditions, we would expect a pulling coefficient of order $10^{-5}$, which is still a very promising number.  In order to limit frequency shifts associated with cavity detuning to the 1~mHz level, we would thus require a cavity that is stable at the 100~Hz level.  While our current cavity exhibits short-term fluctuations in its resonance frequency of order 5~kHz, and slow drifts of several 10’s of kHz over a typical data taking session of a few hours, relatively straightforward technical improvements should enable stabilization of the cavity frequency to the 100~Hz level.  

\section{Conclusion}\label{sec:conclusion}

We have demonstrated the first optical high-precision active atomic frequency reference to date. We have demonstrated a short-term stability that already greatly surpasses that of existing active atomic frequency references (masers), which operate at microwave frequencies. Our system is highly insensitive to drifts in the optical cavity frequency and to fluctuating magnetic fields.  Major systematic frequency shifts have been characterized to the Hz level. 

In the near future, it will be advantageous to extend the pulsed mode of operation demonstrated here to steady-state operation. 
Doing so could dramatically improve the stability of the frequency reference, particularly at short timescales where stability is limited by the finite duration of the pulses and by photon shot noise. For example, if our system were operated in steady-state with a similar output power to what we demonstrate in pulsed mode, and continued to be dominated by photon shot noise out to an averaging time of one second, we would expect to see an improvement in the $\tau = 1$~s Allan variance by a factor of roughly $10^{3}$ compared to pulsed operation with a single 100~ms long pulse each second. 
The rapid improvement arises because the Fourier-limited linewidth would be reduced by 10, and simultaneously 10 times more  photons would be collected, allowing the narrower line to be split more precisely. 




\section*{ACKNOWLEDGMENTS}\label{ack}
  
We acknowledge useful input from N. Poli, A. D. Ludlow, and S. L. Campbell. This work was supported by NSF PFC grant number PHY 1734006, DARPA QuASAR and Extreme Sensing, and NIST.  J.R.K.C. acknowledges financial support from NSF GRFP.

\section*{Appendix I: Experimental setup and data processing}\label{sec:Appendixapparatus}


Our experiment starts by loading about $4\times10^5 \ ^{87}$Sr atoms into a near-magic-wavelength one-dimensional optical lattice created by the intracavity fields of the cavity. The trap depth varies, but nominally we operate with a trap depth of $U_0 = 180~\mu\mathrm{K}\approx 1000E_\mathrm{rec}$. Cooling into the lattice is performed by the narrow linewidth 7.5~kHz transition at 689~nm, following a similar procedure as detailed in \cite{Norcia2017narrow}. Typically, the measured temperature of the atoms in the lattice is around 12~$\mu$K $\approx 65E_\mathrm{rec}$. To prepare the atoms in a near-equal mixture of the two stretched ground states, we optically pump the atoms using a $\pi$-polarized 689~nm laser tuned to the $^{1}$S$_{0} \ F=9/2 \ \to \  ^{3}$P$_{1} \ F^\prime=7/2$ transition.  

\begin{figure}[!htb]
\includegraphics[width=3.375in]{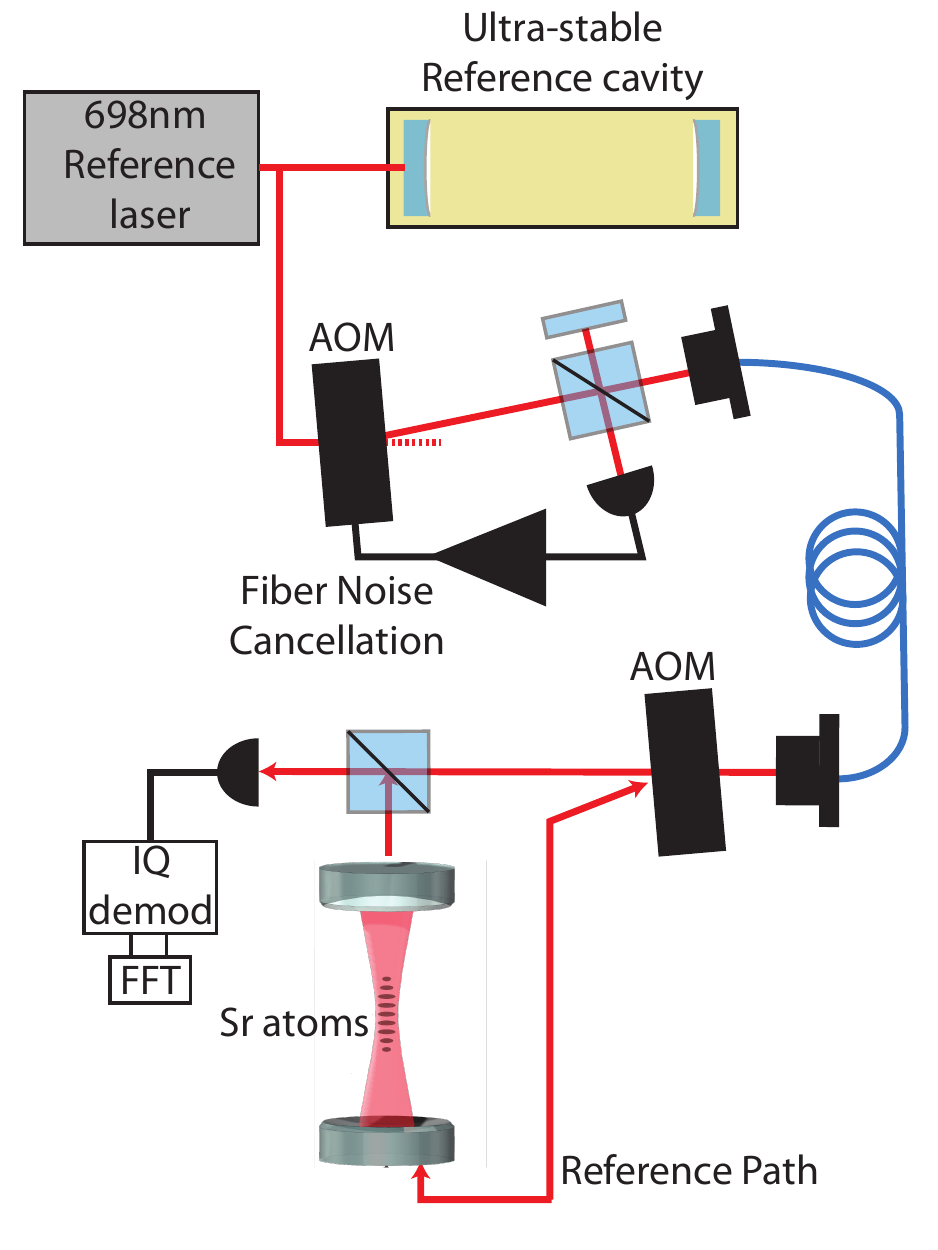}
\caption{Light from a state-of-the-art stabilized cavity laser system is transported by phase stabilized a 75~m optical fiber and is combined with the superradiant pulse light to originate an optical beat-note detected at the photodiode. An IQ demodulation step is performed before data acquisition (FFT). The reference path for the phase stabilization is referenced to the back cavity mirror.}
\label{fig:system}
\end{figure}

The superradiant emission exits the cavity equally through each cavity mirror and it is collected in an optical fiber on one side of the cavity. The total  detection efficiency is $\eta = 0.06$, which accounts for real and effective losses, including the fact that half of the photons exit through the undetected end of the cavity. Light from the reference laser is also coupled to the optical fiber to form an optical beat-note on a fast photodetector, as shown in Fig. \ref{fig:system}. The two fields differ in frequency by approximately  60~MHz, which is within the detection bandwidth of the photodetector.  The observed photo-electron shot noise is approximately 8~dB above the electronic noise floor of the detector. We further mix the signal down using an IQ demodulator to translate the signal close to 40 kHz.  The I and Q outputs are  directly sampled into our data acquisition system. 

Acoustic noise in the 75~m long optical fiber that delivers the reference laser light leads to fluctuations of the reference laser frequency of order 10~kHz, which preclude the observation of a beat-note. We retro-reflect a small fraction of the reference light to actively cancel this noise using the technique demonstrated in ref.~\cite{ma1994delivering}, which reduces contribution of this noise to below at least the 1~Hz level between measurements. Because the cavity can move with respect to the optical table upon which the superradiant pulse is overlapped with the reference laser, it is important to reflect the path-length stabilization light from the cavity mirror opposite to the end of the cavity from which superradiant light is detected, as shown in Fig \ref{fig:system}.  

\section*{Appendix II: Shifts on the clock transition.} \label{sec:AppendixShifts}
Table ~\ref{tab:shifts} summarizes the corrections and uncertainties associated with the measurement of the absolute frequency.  
After accounting for these shifts, the corrected  frequency difference between the superradiant light and the strontium transition frequency measured by the optical lattice clock is $1 \pm 2$~Hz.

\begin{center}
\begin{table}[!htb]
\centering
\begin{tabular}{ |p{4cm}||p{2cm}||p{2cm}|  }
 \hline
 \multicolumn{3}{|c|}{Corrections and Uncertainties} \\
 \hline
 \bf{Effect} & \bf{Value~or Shift (Hz)} & \bf{Uncertainty (Hz)}\\
 \hline
 \hline
 Measured Offset ($f_\mathrm{offset}$)& 0.7 & 0.9  \\
 Cavity Drift, Magic Wavelength &0&0.7\\
 Collisions & 1 & 1\\
 Hyperpolarizability & 0.3 &0.3\\
 Optical lattice clock &0&0.3\\
 2nd order Zeeman & -0.04 & 0.01 \\
 DC Stark Shift & 0 & +0, -0.1 \\
 Differential Blackbody Radiation &0&0.06\\
 \hline
\end{tabular}
\caption{Summary of shifts and uncertainties in the frequency comparison measurements with the optical lattice clock. 
} \label{tab:shifts}
\end{table}
\end{center}

\subsection*{Lattice configuration}
The intracavity lattice is driven by an ECDL laser, whose power is increased by a tapered amplifier.  The lattice light is linearly polarized along the direction of the applied bias magnetic field. This configuration greatly suppresses the vector shift to the clock transition due to the lattice. As we use both stretched states, $m_f = \pm 9/2$ of the clock states ($F=9/2$), the lattice stark shift, in this case equal for both $m_f$ states, for a lattice depth $U$ is $\Delta f_{AC}  = -[\kappa_s + 2\kappa_t F(2F-1)]U$, where $\kappa_s$ and $\kappa_t$ are the differential scalar and tensor polarizabilities respectively
\cite{nicholson2015systematic,westergaard2011lattice}. The magic wavelength is then defined as the one that cancels the scalar and tensor shifts for this configuration. From the fitting presented in the main text, we found the magic wavelength to be 813.4275(2)~nm, limited by the resolution of the wavemeter (calibrated using our narrow-line cooling laser at 689~nm), in good agreement with known results \cite{campbell2017fermi,nicholson2015systematic,Shi2015,LeTargat2013}. The sensitivity to trap depth and detuning from the magic wavelength is described by the fitting parameter $a$ mentioned in the main text. Our measured value of $a$ is $-14.8(3)$~($\mu$Hz/$\mathrm{E_{rec}}$)/MHz, comparable to the previously measured value for the sensitivity of the clock transition to the magic wavelength \cite{campbell2017fermi,Shi2015}

\subsection*{Collisional frequency shifts}
Collisions between atoms in the lattice can lead to frequency shifts of the emitted light that scale with atomic density.  

In order to determine the density-dependent frequency shift, we toggle between measuring the frequency of the superradiant light at an atomic density of $\rho_0 \approx 2\times10^{12}$~atoms/cm$^3$, and a density of approximately $\rho_0/4$.  In order to modulate the density at fixed atom number, we spread out the atoms by adding frequency sidebands on the trapping light at $\pm 1$ free spectral range of the cavity, as demonstrated in \cite{Cox2016}.  As adjacent longitudinal modes of the cavity have opposite symmetry around the center of the cavity (the nodes of one mode corresponds to the antinodes of the other mode), adding the frequency sidebands with the appropriate power ratio results in a shallower axial modulation of the trapping potential at the center of the cavity, while preserving the radial confinement (i.e.~ roughly converting the standing wave lattice into a smooth optical dipole trap.). After loading the atoms in the lattice,  the sidebands are applied to allow the atoms to spread axially across the optical potential. After 30~ms, the sidebands are turned off, restoring the lattice potential and pinning the atoms in place. Cooling is performed for 20~ms to return the atoms to their original temperature. Only 10\% of the atoms are lost during this process, and the final temperature of the atomic sample is unchanged as determined from time of flight measurements. This allows us to investigate the effects of density with minimal change to the atom number, which would lead to additional cavity-mediated pulling effects.

We attribute frequency shifts that occur when the atomic density is toggled at fixed atom number to atomic collisions.  For spin-polarized fermionic samples, these shifts can strongly depend on population inversion \cite{Swallows2012density,martin2013quantum}. In our case, the atomic sample is prepared in two spin projection states, such that even for the fermionic species the s-wave collisions are allowed, and we expect that they are the dominant source of frequency shifts.  This conclusion is supported by measuring the collisions shifts in a fully spin polarized sample and observing a significantly reduced density-dependent frequency shift.

The observed collisional shift between $\rho_0$ and $\rho_0/4$ of $1.7(4)$~Hz, independent of population inversion. 
For the frequency comparison and stability measurements presented in Fig.~\ref{fig:allan}(a)-(c), the density is reduced to roughly $\rho_0/5$ by using a lower atom number, without spreading out the atoms.   In Table ~\ref{tab:shifts} we present a conservative atomic collision shift of $1\pm1$~Hz.  The uncertainty on this number reflects both uncertainty in collisional shifts at a density of $\rho_0$ and the uncertainty in the relative total atom number between when the density shifts were measured and when the absolute frequency measurements were performed.

\subsection*{Other shifts}

Other known sources of frequency shifts that can affect the comparison are estimated to be negligible at the level of our frequency comparison measurement. These include hyperpolarizability, 2nd order Zeeman shifts, DC Stark shifts and  differential blackbody radiation shifts between the clock and superradiant source. These are summarized below and shown in Table ~\ref{tab:shifts}.

As our lattice can be relatively deep compared to usual optical lattice clock experiments, the hyperpolarizability effect that scales quadratically with the lattice intensity could be important. Using the measured coefficient in Ref.~\cite{nicholson2015systematic}, we expect a shift of $0.3(3)$~Hz for the deepest lattice used in this experiment.

The second-order Zeeman shift changes the transition frequency of  both stretched states  by the same amount (-0.233~Hz/G$^2$)$B^2$  \cite{boyd2007nuclear} and does not cancel in the average frequency $f_m$. The half frequency difference $f_d$  between the peaks generated by the  $m_f=\pm9/2$ states provides an excellent estimate of the magnetic field on each trial.  We estimate a shift of -0.04(1)~Hz.

The uncertainty in the optical lattice clock's lattice-induced shift was 0.3~Hz at the time of the frequency comparison. This was due to the 100~MHz uncertainty of the wavemeter used to measure the frequency of the lattice laser.  

Blackbody radiation shifts the transition frequency in both the superradiant light source and in the passive optical clock \cite{nicholson2015systematic,Katori2015,Hinkley2013, Middelmann2012,Safronova2013}.  We measured the temperature of both vacuum chambers to be well within 1~K of each other. However, it was challenging to measure the temperature of the clock chamber very close to the atoms, so we allow for a temperature difference of up to 2~K. Accounting for the variation of the blackbody radiation shift with respect to temperature \cite{nicholson2015systematic} of 0.03~Hz/K, the uncertainty on the differential shift due to blackbody radiation is 0.06~Hz.  We assume that similar characterization of the environment could in-principle be done to establish the absolute magnitude of the shift as was done in \cite{nicholson2015systematic,Hinkley2013} or by placing the atoms in a low temperature environment as was done in \cite{Katori2015}.

Static electric fields can shift the atomic transition due to differential DC polarizability of the excited and ground states. It has been reported that dielectric surfaces under vacuum can accumulate surprising amounts of charge, and lead to electric fields that can cause significant perturbation to the frequency of atomic clocks \cite{Lodewyck2012}. This effect could be relevant in our experiment, due to the ceramic cavity spacer and mirrors that are in close proximity to the atoms. Furthermore, the piezo tubes that control the position of the cavity mirrors are driven with up to 120~V, although they are arranged in a configuration were the electric field generated by the piezos at the cavity center should be approximately canceled. For the data presented in Fig. \ref{fig:allan}(c), the value of the piezo voltage for each lattice configuration can be taken into account for the fitting, as an additional quadratic term on the voltage, without significant changes on the fit quality.

The magnitude of the DC stark shift scales with the square of the electric field at the location of the atoms.  The magnitude of the shift due to stray fields can be determined by applying an additional field.  As the strength of the applied field is scanned, one observes a parabolic frequency shift of the clock transition with a minimum when the applied field just cancels the component of the stray field along the direction of the applied field. This can be done along three different directions to determine the frequency shift due to the total stray field \cite{nicholson2015systematic,Lodewyck2012}.

To generate an applied field, we used a simple electrode placed on the top vacuum chamber viewport, above the horizontally-oriented cavity. We applied a voltage to this electrode relative to the vacuum chamber (which is grounded), and varied this voltage over roughly $\pm 500$~V. We repeated this measurement with different electrode positions on the top viewport in order to vary the direction of the applied field, and observed frequency offsets of the parabolic variation with applied voltage always well below 0.1~Hz,  as determined by comparing the frequency measured at the minimum of the parabola to its value at zero applied voltage. We thus assign a correction and uncertainty associated with DC Stark shifts of -0.05(10)~Hz

\section*{Appendix III: Photon shot noise and reference laser frequency noise aliasing in the superradiant pulsed laser} \label{sec:appendixDE}

The instability in the difference frequency $f_d$ (red points in Fig. \ref{fig:allan}(a)) is dominated by photon shot noise. For a fixed duration of pulse, the photon shot noise limited uncertainty in our frequency measurements should scale as $\delta f \propto 1/\sqrt{M}$, where $M$ is the number of detected photons.  We verify this scaling by attenuating the superradiant emission while holding reference power fixed, and measuring the increase in the short-time Allan deviation of $f_d$.  We find that for a factor of 4 attenuation, the short-time Allan deviation increases by a factor of of 1.9(2), consistent with photon shot noise.  Under the same conditions, the short-time Allan deviation of the average frequencies $f_m$ increased by a factor of 1.4(2), confirming that this quantity was not purely limited by photon shot noise.  As the total detection efficiency is only $\eta = 0.06$, see Appendix I, the photon shot noise represents a more stringent limit than atom shot noise in our setup.

Frequency noise on the reference laser contributes to our measured Allan deviation for the average frequency $f_m$ in a manner that depends both on the noise spectrum of the laser, and on the protocol used to compare its frequency to that of the superradiant source.  For a periodic measurement protocol, as is used here and in typical passive atomic clocks, the contribution to the Allan variance at a time $\tau = n T_c$, where $T_c$ is the cycle time of the experiment, can be quantified as \cite{riehle2006frequency,martin2013thesis}

\begin{equation}
    \sigma^2(nT_c) =  \int_{0}^{\infty} df ~ \frac{2 \sin^4(\pi f n T_c)}{n^2 \sin^2(\pi f T_c)} ~S_{\Delta f}^{1s}(f) |R(f)|^2, \label{eqn:DEdiscrete}
\end{equation}

\noindent Here, $S_{\Delta f}^{1s}(f)$ is the single-sided laser frequency noise power spectral density and $R(f)=\mathcal{F}[r(t)](f)$ quantifies the sensitivity of a single measurement to laser noise at a given frequency, as computed from the Fourier transform of the sensitivity function $r(t)$ described below.  $R(f)$ depends on the manner in which the laser frequency is measured, for example on whether it is measured using a Ramsey sequence, a Rabi sequence, or a beat-note as performed here.  

For long averaging times $\tau$, corresponding to $n \gg 1$, the reference laser's contribution to the Allan variance takes the form 
\begin{equation}
\sigma^2(\tau) = \lim_{n\to\infty} \sigma^2(nT_c) =  \frac{1}{\tau}\sum_{m=0}^{\infty}  S_{\Delta f}^{1s}(m/T_c) |R(m/T_c)|^2. \label{eqn:DEcont}
\end{equation}
which samples discrete frequency components of the reference laser's noise spectrum.  This effect is often referred to as Dick noise aliasing \cite{Dick1987}.

$R(f)$ can be calculated by taking the Fourier transform of a quantity $r(t)$, the time-domain sensitivity function which represents the sensitivity of the frequency measurement procedure to a discrete phase jump at time $t$ \cite{riehle2006frequency,martin2013thesis}.  To estimate $r(t)$ for our measurement sequence, we numerically calculate the sensitivity of our data-analysis procedure by applying phase-jumps to a simulated signal that matches the typical profile and duration of our superradiant pulses.  

\begin{figure}[!htb]
\centering
\includegraphics[width=3.375in]{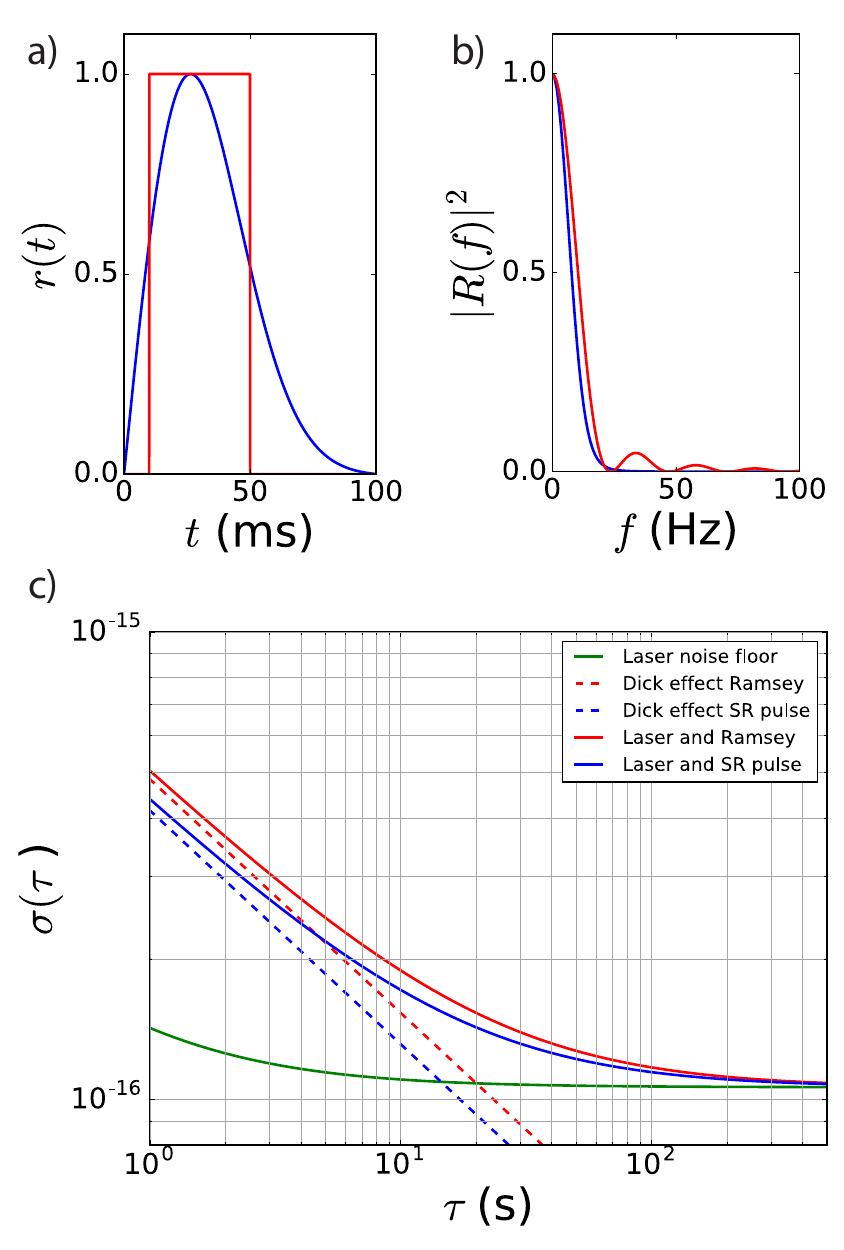}
\caption{Frequency noise aliasing. (a) Sensitivity function $r(t)$ for a 40~ms duration Gaussian superradiant (SR) pulse centered at $t=$15~ms (blue), and for a Ramsey sequence with $T$=40~ms (red). (b) Modulus squared of the normalized Fourier transform of the sensitivity functions shown in (a). (c) Contribution to fractional Allan deviation from reference laser, under different measurement scenarios: Allan deviation of reference laser, assuming no dead-time in measurement (green) \cite{Swallows2012,Bishof_2013},  Allan deviation contribution of reference laser for SR pulse (dashed blue),   quadrature sum of laser noise and frequency aliasing for SR pulse (solid blue), and quadrature sum of laser noise and frequency aliasing for Ramsey sequence (solid red).  }
\label{fig:DEfig}
\end{figure}

The result of this calculation is shown in Fig.~\ref{fig:DEfig}~(a) (blue trace).  Intuitively, this indicates that our measurement protocol is most sensitive to phase jitter that occurs in the middle of the superradiant pulse.  

For comparison, the sensitivity function for a Ramsey sequence of duration 40~ms that begins at $t=$10~ms is shown as the red trace in Fig.~\ref{fig:DEfig}~(a) \cite{Santarelli1998}.  Unlike our heterodyne frequency measurement, the Ramsey sequence has equal sensitivity to a phase jump any time during its duration.  

The frequency-domain sensitivity function $|R(f)|^2$ for both the heterodyne detection and Ramsey sequence are shown in Fig.~\ref{fig:DEfig}~(b) (blue and red traces, respectively). The key result here is that $R(f)$ falls off at a frequency given roughly by the inverse of the duration of a single measurement sequence.  Shorter superradiant pulses or Ramsey sequences lead to appreciable sensitivity at higher frequencies, and an increase in the area under the frequency-domain sensitivity curve.  

Fig.~\ref{fig:DEfig}~(c) illustrates the effect of this increased sensitivity due to short measurement sequences.  For a continuous measurement, the Allan deviation of the reference laser is known, and is represented by the green line. Using the sensitivity functions for our superradiant pulses and for 40~ms long Ramsey sequences, we compute the blue and red lines, respectively.  

From the above, we estimate that at 1~s, the reference laser's frequency noise contributes $\sigma(1~\mathrm{s}) = 4.4\times 10^{-16}$ to the total Allan deviation. For the data with the best observed short term Allan deviations (blue points in Fig. \ref{fig:allan}(a) in the main text), the difference between the Allan deviation for the frequency average, $f_m$, and difference, $f_d$, can be explained by the reference laser's frequency noise.  

\bibliography{ThompsonLab.bib}

\end{document}